# Accurate Measurement of Propagation Delay in a Multi-Span Optical Link

Michael H. Eiselt
*ADVA*
Meiningen, Germany
meiselt@advaoptical.com

Florian Azendorf
*ADVA*
Meiningen, Germany
fazendorf@advaoptical.com

*Abstract*—The principle of Correlation Optical Time Domain Reflectometry (C-OTDR) is proposed to accurately measure the propagation delay over a multi-span optical fiber link. The delay of the transmission fiber is measured in the reflective mode, while uni-directional node components are measured in a transmissive mode. Delimiting reflectors are required between the sections for accurate demarcation.

## I. Introduction

Optical fiber delay and delay asymmetry have recently attracted increasing attention due to the impact on synchronization applications, e.g. in 5G front haul links. Especially for synchronization applications based on the IEEE 1588v2 standard, or precision time protocol (PTP), a difference between the signal propagation times from the master clock to the slave clock and vice versa results in a timing error of the slave clock, as the protocol assumes equal propagation delays. Similarly, the common public radio interface (CPRI)[1], used for mobile front haul from the base band unit to the remote radio head, allows for a maximum asymmetry of 8 ns between the two propagation directions. An even stricter requirement on the asymmetry between multiple links applies for the transmission of radio signals to the elements of a phase array antenna [2]. Here, only few picoseconds of asymmetry are acceptable.

For all these applications, differences of the fiber propagation delay, due to differences in length or refractive index, must be determined upon installation. However, temperature changes in the environment also lead to a change in the propagation time over each fiber and might also impact the propagation delay difference between fibers. In [3], the propagation delay of a 8.5-km fiber link had been monitored over time, and a difference of more than 12 ps was observed between the fibers in a cable. The overall delay variation was around 800 ps during this time. This points to the necessity of monitoring the fiber delay during operation.

A method for the monitoring of the propagation delay of up to 100 km fiber with an error of a few picoseconds was reported in [4], [5]. The correlation optical time-domain reflectometry (C-OTDR) method relies on the reflection of a probe signal at the fiber end and backward propagation of the reflected signal over the fiber. In an optical link, however, some components, like an optical amplifier, contain isolators and therefore permit only uni-directional transmission. In this paper, an extension of the C-OTDR method is proposed, enabling an accurate measurement also for uni-directional link components.

## II. Principle of the Correlation OTDR

In the well-known OTDR technique, an optical pulse of several nanoseconds up to a few microseconds' duration is transmitted into an optical fiber under test, and the backscattered and reflected power is recorded as a function of time. The round-trip time of the received signal is a measure of the reflection location in the fiber. Due to the ns-length of the pulse and the propagation time of approximately 5 ns/m in standard fiber, the resolution of this method is on the order of meters. Using shorter pulses with a better time and space resolution would reduce the signal-to-noise ratio (SNR) of the reflected signal and require an increased number of averages, resulting in a longer measurement time. In the C-OTDR technique, a bit burst at a higher data rate, e.g. 10 Gbit/s, is transmitted into the fiber, and the reflected signal is cross-correlated with the transmitted bit pattern. Due to the high bit rate of the burst, the temporal resolution is approximately 100 ps, while a similar SNR is obtained as with a pulse of the whole burst length. Further signal processing can improve the temporal resolution to a few picoseconds [4].

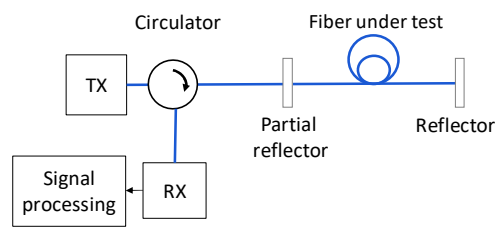

Fig. 1. Schematic setup of C-OTDR measurement. TX: transmitter, RX: receiver.

A schematic setup of the C-OTDR measurement is shown in Fig. 1. The probing bit sequence is inserted, via a circulator, into the fiber under test. To eliminate the impact of the delay in the jumper cables and to obtain a precise value of the fiber under test round-trip delay, the input end of the fiber is defined by a partial reflector, which reflects part of the probe signal back into the circulator. Another part of the probe signal propagates along the fiber and is reflected at intermediate reflection points (e.g. connectors) and another delimiting reflector provided at the fiber end. A receiver at the third port of the circulator records the

This work has received funding from the European Union´s Horizon 2020 research and innovation programme under grant agreement No 762055 (BlueSpace project).

time evolution of the reflected power, and in a signal processing step a cross-correlation (CC) between the received signal and the transmitted bit sequence is carried out. Peaks in this cross-correlation determine the round-trip time with a resolution of the bit duration. By oversampling the received signal, i.e. sampling at a higher rate than the burst bit rate, the shape of the signal is imaged in the shape of the CC function. A raised-cosine function, corresponding to the expected pulse shape, is fitted to the CC function, and the center of this function is used to determine the round-trip time with an accuracy of a fraction of the bit duration [3], [4].

### III. Measurement of Link Delay

Fig. 2 shows the schematic of a multi-span optical link. Transmitter (TX) and receiver (RX) are connected via optical fiber spans and optical nodes, the latter, for example, containing optical amplifiers. To measure the propagation delay from TX to RX, a signal with a timing information could be transmitted by the TX and received by the RX. The difference of transmit time and receive time, as monitored by clocks at the TX and RX, respectively, corresponds to the propagation time. This requires, however, synchronized clocks on both sides of the link. A time offset between the two clocks would lead to an error in the delay calculation. To synchronize the two clocks, e.g. using PTP, however, knowledge about the delay – or at least the delay asymmetry between the two directions – would be required first. It is therefore necessary to use a single-ended measurement, where transmit and receive time can be derived from the same clock.

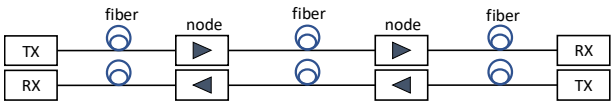

Fig. 2. Bi-directional optical multi-span link.

The reflective C-OTDR method can be used to measure the propagation delay of the fiber sections of the link in a reflective way. The node sections, however, might not be transparent for the reflected signal, as, for instance, optical amplifiers contain isolators. Here, the C-OTDR method can be applied to the signal transmitted through the node in forward direction to measure the delay time. To delineate the sections, delimiting reflectors are required, which can also be used for calibration of the cable length of the measurement setup for the measurement in forward direction.

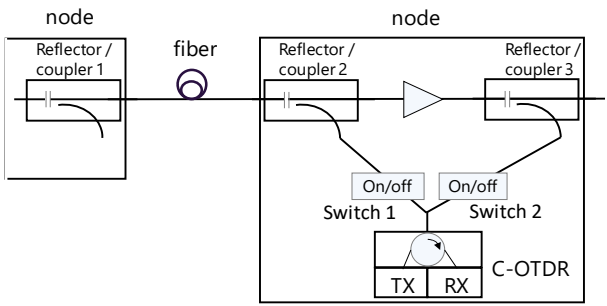

Fig. 3. Node setup for link delay measurement. Only part of the left neighbor node is shown.

Fig. 3 depicts a node setup with a C-OTDR, which is connected to measure the fiber delay to the neighboring node as well as the transition time of a signal through the node. A combination of couplers and switches enables the alternate measurement of fiber and node components. The individual delay contributions can be measured in the following steps:

1. Turn on only Switch 1 and transmit probe signal. Receive signal from near end reflector 2 and from far end reflector 1. Calculate the round-trip fiber delay of the attached fiber span as the difference between both. Also record round-trip delay to reflector 2.

2. Turn on only Switch 2 and transmit probe signal. Receive signal from reflector 3. Record round-trip delay to reflector 3.

3. Turn on Switch 1 and Switch 2 and transmit probe signal. Receive signal through the uni-directional device. Calculate the single-pass delay between reflectors 2 and 3 by subtracting half the round-trip delays to these reflectors, as recorded in steps 1 and 2, from the measured single-pass delay.

Using this method, the overall link delay can be obtained by combining all measured fiber delays and all delays of the uni-directional devices. Each individual delay is defined between the delimiting reflectors.

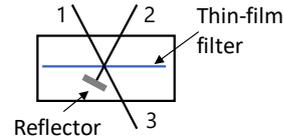

Fig. 4. Implementation option for delimiting reflector.

While the reflector/coupler combination shown in Fig. 3 is a possible realization of the delimiting reflector, it also introduces loss to the data signal travelling over the fiber link. An alternative realization, depicted in Fig. 4, uses a thin-film filter (TFF), which fully reflects data signal wavelengths and partially reflects the probe wavelength. One TFF port is terminated by a strong reflector. Data signals can thus travel from port 1 to port 2 of the device with suffering only minor loss. A probe signal entering the delimiting reflector at port 3 passes through the TFF to port 1 with a small (~ 1 dB) attenuation. Likewise, a probe signal entering at port 1, passes to port 3 with the small attenuation. A part (~ -7 dB) of the probe signal entering at port 3 is reflected at the TFF towards the reflector and then passes the TFF again to port 2. A signal entering at port 2, like the probe signal having travelled over the fiber, mainly passes the TFF, is reflected at the reflector, and is reflected back to port 2. The delimiting reflector can therefore be used on both sides of the node, like the reflector/coupler combinations in Fig. 3. As the reference plane for the delimiting reflector is the reflection point on the TFF, the distance between TFF and strong reflector must be calibrated for an exact measurement.

### IV. Measurement Accuracy

The delay measurement of a 100-km fiber span has been reported in [5]. While the position accuracy, based on the precise determination of the cross-correlation peak, was on the order of a few picoseconds, the accuracy of the delay measurement mainly depends on the accuracy of the clock frequency used as time base for recording the received signal. For 100 km of fiber, the propagation delay is approximately 500 µs. A delay accuracy of better than 10 ps requires a relative clock accuracy of 20 ppb (20 x $10^{-9}$). This high accuracy can be achieved, when the clock is locked to the timing signal from a Global Navigation Satellite System (GNSS). The typical accuracy of a low-cost oscillator is on the order of 100 ppm ($100 \times 10^{-6}$) and would, when used in this application, therefore yield a timing error of 50 ns. In most applications, however, the absolute delay is of less relevance than the delay difference between the opposite link directions. In systems, where the signals in both directions travel over fibers in the same cable and use similar node equipment, the propagation length difference is typically not more than a few tens of meters, corresponding to a few 100 ns delay difference. If both directions of a fiber span are measured with the same clock, the accuracy, even when using the low-cost oscillator, is better than a few tens of picoseconds for a 100-km fiber link. To achieve this accuracy, it is necessary to measure the two link directions from the same node within a short time interval, e.g. by switching the C-OTDR equipment between the two fibers. Due to the typical temperature coefficient of approximately 7 ppm/K [3], [6], the fiber delay of a 100-km link changes by 35 ps for a small temperature change by one hundredth of a degree Celsius. If a very high level of accuracy is required, e.g. to maintain a tight synchronization between master and slave clock, continuous monitoring of the link delay is required to detect changes in the delay asymmetry due to temperature variations.

## V. SUMMARY

The use of a correlation OTDR method was proposed to measure and monitor the propagation delay over an optical fiber link. While fiber sections can be measured single-ended in the reflective mode of operation, for uni-direction node components a transmissive method, using the same equipment, is applied. To achieve accurate results, delimiting reflectors for the probe signal are required as defined demarcation points between the link sections. If only the asymmetric delay is of importance for the system application (e.g. for PTP synchronization), low cost oscillators are sufficient to maintain an accuracy of a few tens of picoseconds.